\documentclass[12pt]{article}
\usepackage{amssymb,amsmath,latexsym,graphicx, mathrsfs, lscape, amsthm, bm,  hyperref}

\hypersetup{
    colorlinks=true,
    linkcolor=blue,
    citecolor=red,      
    urlcolor=cyan,
}

\numberwithin{equation}{section}

\begin{document}

\begin{titlepage}

    \thispagestyle{empty}

    \vspace{82pt}
    \begin{center}
        { \LARGE{\bf Schwarzschild Instanton in Emergent Gravity}}

        \vspace{10pt}

        {\large{\bf Sumanto Chanda $^\clubsuit$, Partha Guha$^\clubsuit$ and \ Raju Roychowdhury$^{\spadesuit}$}}

        \vspace{10pt}

        {$\clubsuit$ \it S.N. Bose National Centre for Basic Sciences\\
        JD Block, Sector III, Salt Lake, Kolkata 700098, India \\
        \texttt{sumanto12@boson.bose.res.in, partha@bose.res.in}}

        \vspace{10pt}
       
        {$\spadesuit$ \it Instituto de Fisica, Universidade de Sao Paulo,\\
                C. Postal 66318, 05314-970 Sao Paulo, SP, Brazil\\
        \texttt{raju.roychowdhury@gmail.com, raju@if.usp.br}}

       \end{center}

\abstract{In the bottom-up approach of emergent gravity we attempt to find symplectic gauge
fields emerging from Euclidean Schwarzschild instanton, which is studied as electromagnetism 
defined on the symplectic space $(M,\omega)$. Geometrical engineering with the 
emergent metric sets up the Seiberg Witten map between commutative and non-commutative gauge fields, preparing the 
ground for the evaluation of topological invariants in terms of the underlying gauge theory quantities.}

\bigskip 



\vspace{-0.25cm}
\tableofcontents

\end{titlepage}

\setcounter{page}{1}
\section{Introduction}

Gravity and supergravity are gauge theories, just like Yang-Mills theory, implying that gravitational 
instantons play the same role as Yang-Mills instantons. Instantons (For a set of self-contained lectures look for \cite{instantlecture}) are non singular solutions of classical equations in 4-dimensional Euclidean space and a useful tool to study low-dimensional sigma 
models and supersymmetric QCD. Since instantons are non-perturbative objects they play an important role 
in defining the vacuum structure of QCD and it was found by Belavin, Polyakov, Schwarz and Tyupkin \cite{bpst} 
that is why in literature it is known after BPST instantons. Their role is listed as:
\begin{enumerate}
\item providing stationary phase points in path integrals for amplitude to tunnel between topologically 
distinct field configurations \cite{gibhawk1}
\item possibly playing a role in quark confinement \cite{adg, ccg} (for more details see \cite{thbr, pvb, ss}) that leads baryons to decay into leptons and asymptotic freedom of QCD
\item contributing to the anomalous divergence of the axial vector current \cite{gthooft}
\end{enumerate}
Gravitational instantons are non singular complete positive definite metrics satisfying the classical 
vaccum Einstein equations or the Einstein equations with $\Lambda$ (a Lagrange multiplier for 4-volume or 
the outcome of certain supergravity Lagrangians \cite{hawk1}). In Euclidean quantum gravity, they are stationary 
phase metrics in the path integrals for the partition functions, $Z$ \cite{gibhawk1}, of the thermal and volume canonical 
ensembles \cite{hawk1, gibhawk2}. In these cases the instanton action dominates the contribution to $- \log Z$. This 
action is related to the areas of the bolts and to the nut charges and potentials. Nuts and bolts 
exhibit a symmetry analogous to duality invariance in electromagnetism. Bolts are analogous to 
``electric" type mass monopoles and nuts to gravitational dyons with a real electric type mass-monopole and an imaginary ``magnetic" type mass-monopole. The appearance of magnetic monopole induces Dirac string-like singularity into the metric
which can be further removed by appropriate identifications and changes in the topology of the four manifold. So Nuts cannot
occur in the classical regimes without some quantum fluctuations of the background contrary to the appearance of bolts. 
This implies that the bolts have an intrinsic gravitational entropy equal to one quarter the sum of their areas. This generalises the 
results obtained for black holes and cosmological event horizons \cite{gibhawk2, hawk2, gibhawk3}. 

The Euclidean Schwarzschild solution is a canonical example of a gravitational instanton exhibiting 
one parameter continuous symmetry group as opposed to two parameter continuous  symmetry group exhibited by
almost  all other known gravitational instantons. It is an well known asymptotically flat (AF) gravitational instanton alongside 
the Euclidean Kerr and flat space $S \times \mathbb{R}^3$ which is  a trivial examle representing the class \cite{lap}. Flat 
space $E^4$ is also known to be the unique asymptotically Euclidean gravitational instanton. It was an unproven 
conjecture that Euclidean Schwarzschild and Kerr are the only non-trivial  AF gravitational instantons 
besides flat space, due to some blackhole uniqueness theorems which was later proven to be false by 
Chen and Teo \cite{chenteo}. 

It is worth paying attention to the fact that the thermal nature of black hole emission can be related 
directly to the properties of Euclidean Schwarzschild solution ala Hawking. In the Euclidean 
approach to quantum field theory one attempts to define quantities on a ``Euclidean section" and 
then obtain the physical spacetime quantities by analytic continuation. Particularly, the
Feynman propagator for a field on spacetime is obtained by analytically continuing the Green's 
function on Euclidean section. Thus one is naturally led to study and examine the salient features
of Euclidean Schwarzschild solution. \\ \\
Mathematically the Euclidean Schwarzschild $4$-manifold $M$ 
is a complete solution to the Euclidean Einstein's equations with zero
cosmological constant $\Lambda$, and has the non-trivial topology 
$M\cong \mathbb{R}^2\times S^2$. In other words it
is a Ricci flat manifold. It is not a self-dual solution
 (e.g. the Taub-NUT metric or the Eguchi-Hanson metric) although classified as an AF type gravitational instaton.
We have a particularly nice form of the metric $g$ on a dense open subset 
$(\mathbb{R}^2\setminus \{ O\})\times S^2\subset M\cong \mathbb{R}^2\times S^2$ of the
Euclidean Schwarzschild manifold. It is convenient to use polar
coordinates 
$(r,\tau)$ on $\mathbb{R}^2\setminus\{O\}$ in the range $r\in (2m,\infty)$ and 
$\tau\in [0,8\pi m)$, where $m>0$ is a fixed constant related to the mass of the black hole.
The metric then takes the form 

\vspace{-0.25cm}
\[ d s^2=\left( 1-\frac{2m}{r}\right) d \tau^2+
\left( 1-\frac{2m}{r}\right)^{-1} d r^2+r^2 d\Omega^2,\]
where $ d\Omega^2$ stands for the line element of the unit round $S^2$. 
In  spherical coordinates $\Theta\in (0,\pi)$ and $\phi\in [0,2\pi)$ it is 
$$ d\Omega^2= d\Theta^2+\sin^2\Theta\: d\phi^2$$ on the open coordinate
chart 
$(S^2\setminus(\{S\}\cup\{ N \} ))\subset S^2$.
Consequently the above metric takes the following form on the open, dense 
coordinate chart $U:=(\mathbb{R}^2\setminus\{O\})\times (S^2\setminus(\{S\}\cup\{N\}))\subset 
M\cong \mathbb{R}^2\times S^2$:

\vspace{-0.75cm}
\begin{eqnarray}  d s^2=\left( 1-\frac{2m}{r}\right) d\tau^2+
\left( 1-\frac{2m}{r}\right)^{-1} d r^2+r^2 ( d\Theta^2+\sin^2\Theta
 d\phi^2).
\label{metric}\end{eqnarray}

\vspace{-0.25cm}
Despite the apparent singularity of the metric at the origin $ O\in\mathbb{R}^2$,
it can be extended analytically to the whole
$\mathbb{R}^2\times S^2$ as demonstrated in Wald \cite{wald}. The $U(1)$ action defined by $\tau\mapsto \tau + 4m\lambda$ for 
$e^{i\lambda} \in U(1)$ leaves this metric invariant, and thus defines the 
Killing vector field 

\vspace{-0.25cm}
\[ X:={1\over 4m}{\partial\over\partial \tau},\] 

which (together with the $U(1)$ action itself) clearly 
extends to a Killing field on the whole Euclidean Schwarzschild manifold,
which we will denote by $X$. Now consider the differential $1$-form $\xi :=g(X,\:\cdot\:)$ 
dual to $X$. In our coordinate chart $U$ it takes the form 

\vspace{-0.25cm}
\[ \xi={1\over 4m}\left(1-{2m\over r}\right) d\tau .\]

General considerations about Killing's equations on a Ricci flat manifold yield that $ d\xi$ is a 
harmonic $2$-form, which on a complete manifold is equivalent to saying that it is closed and co-closed 
and thus harmonic.

The correspondence between noncommutative (NC) U(1) gauge theory and gravity has gained 
much attention in the context of emergent gravity \cite{review1,review2,review4,hsy-jpcs12}. Current research in the field of 
instantons \cite{salizzoni,yangsalizzoni} reveals that the 
gravitational instantons in Einstein gravity are equivalent to $U(1)$ instantons in NC gauge theory. 
In other words, the self-dual electromagnetism on NC spacetime is equivalent to self-dual Einstein 
gravity \cite{hsyangepl09}. This implies that gravity can emerge from electromagnetism defined in 
NC spacetime. The relation between Yang-Mills instantons and gravitational instantons are further 
understood in \cite{ohparkhsyang} where it was shown that every gravitational insatantons are 
$SU(2)$ Yang-Mills instantons on a Ricci-flat four manifold but the reverse is not necessarily true. 
Gravitational instantons satisfy the same self-dual equations of $SU(2)$ Yang-Mills instantons. 
The gravitational instanton which is a solution of (anti) self-dual gravity emerges either from $SU
(2)_L$ or $SU(2)_R$ Yang-Mills instanton sector. The (anti) self-dual gauge fields constructed 
from Yang-Mills instanton generate (anti) self-dual gravity. In \cite{ohparkhsyang} the result was 
further extended  to include general Einstein manifolds \cite{ohhsyang}: all Einstein manifolds 
with or without cosmological constant are Yang-Mills instantons in $O(4)=SU(2)_L\times SU(2)_R
$ gauge theory but the reverse is not true. In fact they arise as a sum of instantons coming 
both from $SU(2)_L$ instanton and $SU(2)_R$ anti-instanton. This may explain the stability of 
the four dimensional Einstein manifold compared to the five dimensional Kaluza-Klein vacuum.

In this note we deal with a specific example of an Einstein manifold: the Euclidean Schwarzschild 
black hole. It is an Einstein manifold which is Ricci flat. It is argued that this geometry is a sum of 
both $SU(2)_L$ and $SU(2)_R$ instanton. It was discussed in \cite{ohparkhsyang} that
the Euclidean solution outside of the (anti)self-dual gravity is a combination of both $SU(2)_L$ and $SU(2)_R$ 
Yang-Mills instanton. 
Following the bottom-up approach of Emergent Gravity \cite{Lee:2012px}, we construct vector 
fields from the Euclidean Schwarzschild instanton and calculate the equations of motion and 
Jacobi identity. Using the Seiberg Witten map we find the symplectic field strength and check the 
absence of self-duality for Euclidean Schwarzschild. We explicitly show the Ricci flatness and 
shed light on the vacuum Einstein solution as is evident from the energy momentum tensor  that 
can be computed exactly exploiting the relation between spin connections and structure constants 
for the Schwarzschild solution. We further study their geometric properties by calculating the 
topological invariants of the $U(1)$ gauge fields \cite{Lee:2012rb} derived from emergent 
Schwarzschild metric. 

The paper is organized as follows:  In section 2. we review the standard results of the bottom-up 
formulation of emergent gravity \cite{Lee:2012px}, In section 3. we introduce the euclidean 
Schwarzschild solution, we make a wise choice for the Darboux coordinates in which we write the 
corresponding metric, then we obtain the set of symplectic $U(1)$ gauge fields and derive the 
corresponding vector fields and check the Jacobi identity for the Poisson  and Lie algebra. Next 
we realize the Seiberg Witten map between ordinary and NC gauge fields and find that the 
solution is neither self-dual nor anti self-dual. In the next section, from the set of tetrads 
we obtain the spin-connections and the curvature components 
and from that we get the Ricci tensor and obtain Ricci flatness for the metric. Ricci flatness 
condition also translates into a vacuum solution. In the penultimate section,  we compute the bulk 
and boundary contribution to the topological invariants  namely Euler characteristics and the 
Hirzebruch signature complex. Here we also obtain $SU(2)_\pm$ gauge fields for emergent 
Schwarzschild instanton and reconfirm the fact that both the gauge fields make an equal contribution 
to the overall Euler invariant or the signature. Thus emergent Scwarzschild solution can be seen as the 
sum of $SU(2)_{L}$ instantons and $SU(2)_{R}$ anti-instantons, thus explaining the generic 
feature of stability for a Ricci-flat manifold like the one we dealt with. We conclude with some 
comments and future directions. The appendices contain some details of the computations, 
namely some identities from differential geometry that has been used, also the t'Hooft matrices 
and the full set of $SU(2)_\pm$ gauge fields in matrix notation.

\section{Review of  Emergent Gravity formulation  in bottom-up}

The mathematical tool to quantize the dynamical system \cite{abraham} is to specify the Poisson 
structure $\theta$ such that
$$\theta={1\over 2}\sum^N_{A,B=1}\theta^{AB}{\partial \over {\partial x^A}}\wedge {\partial \over {\partial x^B}} \in \Gamma (\wedge^2 TM),$$
and then the differentiable manifold $M$ endowed with $\theta$ describes a Poisson manifold $(M,\theta)$. The 
Poisson structure defines an ${\bf R}$-bilinear antisymmetric operation $\{,\}_\theta$: $C^\infty (M)\times C^\infty (M)\to C^\infty(M)$ 
$$(f,g)\mapsto \{f,g\}_\theta=\langle\theta, df \otimes dg\rangle =\theta^{AB}(x) {\partial f(x)\over {\partial x^A}}{\partial g(x)\over {\partial x^B}},$$
and the Poisson bracket satisfy the Leibniz rule and Jacobi identity as follows:
\[ \begin{split}
& \{f,gh\}_\theta=g\{f,h\}_\theta+\{f,g\}_\theta h,\\
& \{f,\{g,h\}_\theta\}_\theta+\{g,\{h,f\}_\theta\}_\theta+\{h,\{f,g\}_\theta\}_\theta=0,
\end{split} \]
$\forall f,g,h \in C^\infty(M)$. The Poisson structure $\theta$ reduces to symplectic structure when it is nondegerate. \\ \\
The application of Darboux theorem or Moser lemma\cite{abraham} of symplectic geometry to electromagnetism 
defined on the symplectic space gives rise to an equivalence principle. An arbitrary 
deformation of symplectic deformation can not be distinguishable locally from canonical form. The 
electromagnetism on symplectic spacetime can be a theory of gravity\cite{hsyangjhep09}: Starting with symplectic 
form $\omega_0=B$, the deformation of $\omega_0$ generate dynamical gauge fields such that $\omega_1=B+F$, where 
$F=dA$. It is always possible to eliminate $F$ by a suitable coordinate transformation as far as the 
2-form $B$ is closed and nondegenerate because in this case the gauge symmetry becomes a spacetime 
symmetry rather than an internal symmetry. This very fact indeed paves the way for a connection 
between NC gauge fields and spacetime geometry. \\ \\
For a given Poisson algebra $(C^\infty(M),\{,\}_\theta)$, there is a natural map $C^\infty(M)\to TM:f \mapsto X_f$ 
between smooth functions in $C^\infty(M)$ and vector fields in $TM$ such that
\begin{equation}
\label{xf} X_f(g(y)) \equiv \{g,f\}_\theta (y)=\left ( \theta^{\mu\nu}{\partial f(y)\over {\partial y^\nu}}{\partial \over {\partial y^\mu}}\right )g(y),
\end{equation}
for any $g\in C^\infty(M)$. This means that we can obtain a vector field $X_f=X^\mu_f\partial_\mu \in \Gamma(TM_y)$ from a 
smooth function $f\in C^\infty(M)$ defined at $y\in M$ where $X^\mu_f(y)=\theta^{\mu\nu} {\partial f(y)\over {\partial y^\nu}}$. As long as $\theta$ is a Poisson 
structure of $M$, the above formula (\ref{xf}) between Hamiltonian function $f$ and Hamiltonian vector 
field $X_f$ is a Lie algebra homomorphism in the sense that
\begin{equation}
\label{xfg} X_{\{f,g\}_\theta}=-[X_f,X_g],
\end{equation}
where the right hand side is a Lie bracket between Hamiltonian vector fields. \\

From the above arguments, $U(1)$ gauge fields on a symplectic manifold $(M,B=\theta^{-1})$ can be 
transformed into a set of smooth functions
\begin{equation}
\label{sgf}
\begin{split}
\{D_\mu(y)\in C^\infty (M) | D_\mu(y) &\equiv B_{\mu\nu} x^\nu (y)=B_{\mu\nu}y^\nu +\widehat{A}_\mu(y),~\mu,\nu=1,\cdots,2n\} \\
\text{where } \qquad x^\mu(y) &\equiv y^\mu+\theta^{\mu\nu} \widehat{A}_\nu(y) \in C^\infty(M)
\end{split}
\end{equation}
After the map (\ref{xf}) is applied, we obtain Lie algebra homomorphism (\ref{xfg}) between the Poisson 
algebra $(C^\infty(M),\{,\}_\theta)$and the Lie algebra $(\Gamma(TM),[,])$ of vector fields defined by
$$\{ V_\mu =V^a_\mu\partial_a \in \Gamma(TM)|V_\mu(f)(y)\equiv \{D_\mu(y),f(y)\}_\theta,~a=1,\cdots,2n\},$$
for any $f\in C^\infty(M)$. The vector fields $V_\mu=V^a_\mu(y){\partial \over {\partial y^a}}\in \Gamma(TM_y)$ take values in the Lie algebra of 
volume preserving diffeomorphisms ($\partial_a V^a_\mu=0$). However, it can be shown that the vector fields 
$V_\mu\in \Gamma(TM)$ are related to the orthonormal frames (vielbeins) $E_\mu$ by $V_\mu=\lambda E_\mu$ where $\lambda^2={\rm det} V^a_\mu$. 
The metric is constructed from these vector fields:
$$ds^2=\delta_{\mu\nu}E^\mu \otimes E^\nu =\lambda^2 \delta_{\mu\nu}V_a^\mu V_a^\nu dy^a\otimes dy^b,$$
where $E^\mu=\lambda V^\mu\in\Gamma(T^*M)$ are dual one-forms. \\

The electromagnetic fields in the symplectic spacetime $(M,B)$ manifest themselves only as a 
deformation of symplectic structure such that the resulting symplectic spacetime is described by 
$(M,B+F)$ where $F=dA=L_{X}B$. This is equivalent to a deformation of frame bundle over 
spacetime manifold $M$: $\partial_\mu\to E_\mu=E_\mu^a(y)\partial_a$, or, in terms of dual frames, $dy^\mu\to E^\mu=E^\mu_a(y)dy^a$. 
$$ds^2=\delta_{\mu\nu}dy^\mu \otimes dy^\nu \to ds^2=\delta_{\mu\nu}E^\mu\otimes E^\nu.$$
We can show the emergence of gravity from the gauge fields starting with the action:
$$S_p={1\over {4g^2_{YM}}}\int d^{2n}y\{D_\mu(y),D_\nu(y)\}_\theta \{D^\mu(y),D^\nu(y)\}_\theta.$$
where $g_{YM}$ is s $2n$-dimensional gauge coupling constant.  Note that
$$\{D_\mu(y),D_\nu(y)\}_\theta = -B_{\mu\nu}+\partial_\mu\widehat{A}_\nu(y)-\partial_\nu\widehat{A}_\mu(y)
+\{\widehat{A}_\mu(y),\widehat{A}_\nu(y)\}_\theta = -B_{\mu\nu}+\widehat{F}_{\mu\nu}(y),$$
\begin{equation}
\label{dd} \{D_\mu(y),D_\nu(y)\}_\theta = -B_{\mu\nu}+\widehat{F}_{\mu\nu}(y),
\end{equation}
$$\text{and } \qquad \{D_\mu(y),\{D_\nu(y),D_\lambda(y)\}_\theta\}_\theta = \partial_\mu\widehat{F}_{\nu\lambda}(y)+
\{\widehat{A}_\mu(y),\widehat{F}_{\nu\lambda}(y)\}_\theta = \widehat{D}_\mu\widehat{F}_{\nu\lambda}(y),$$
\begin{equation}
\label{ddd} \{D_\mu(y),\{D_\nu(y),D_\lambda(y)\}_\theta\}_\theta = \widehat{D}_\mu\widehat{F}_{\nu\lambda}(y).
\end{equation}
By identifying $f(y)=D_\mu(y)$ and $g(y)=D_\nu(y)$ with the relation of (\ref{dd}), the Lie algebra homomorphism (\ref{xfg}) leads to the following identity
$$X_{\widehat{F}_{\mu\nu}}=[V_\mu,V_\nu],$$
where $V_\mu\equiv X_{D_\mu}$ and $V_\nu\equiv X_{D_\nu}$ and using (\ref{ddd}) we have
$$X_{\widehat{D}_\mu\widehat{F}_{\nu\lambda}}=[V_\mu,[V_\nu,V_\lambda]].$$
Thus the equation of motion and the Jacobi identity can be written as
\[ \begin{split}
\{D^\mu,\{D_\mu,D_\nu\}_\theta\}_\theta &= \widehat{D}^\mu\widehat{F}_{\mu\nu}=0,\\
\{D_{[\mu},\{D_\nu,D_{\lambda]}\}_\theta\}_\theta &= \widehat{D}_{[\mu}\widehat{F}_{\nu\lambda]}=0.
\end{split} \]
With the help of the above formula we have the following insightful correspondence
\[ \begin{split}
\widehat{D}_{[\mu}\widehat{F}_{\nu\lambda]}=0 \quad &\Leftrightarrow \quad [V_{[\mu},[V_\nu,V_{\lambda]}]]=0,\\
\widehat{D}^\mu\widehat{F}_{\mu\nu}=0 \quad &\Leftrightarrow \quad [V^\mu,[V_\mu,V_\nu]]=0.
\end{split} \]
These relations reduce to the Einstein field equations and the first Bianchi identity for the Riemann tensor
\[ \begin{split}
[V^\mu,[V_\mu,V_\nu]]=0 \quad &\Leftrightarrow \quad R_{\mu\nu}-{1\over 2}g_{\mu\nu}R={8\pi G\over c^4}T_{\mu\nu},\\
[V_{[\mu},[V_\nu,V_{\lambda]}]]=0 \quad &\Leftrightarrow \quad R_{[\mu\nu\lambda]\rho}=0.
\end{split} \]
where the 2nd equation above implies that individual Riemann curvature components can be
\begin{equation}
\label{bcurv} [V_{\mu},[V_\nu,V_{\lambda}]] = {R_{ \mu \nu \lambda}}^ \rho V_\rho
\end{equation}
This equation will be of relevance to us later, in the next section as we shall see.

\section{Gauge Fields from Euclidean Schwarzschild}
 
The Euclidean Schwarzschild metric is given by:
\begin{equation}
\label{esmetric} ds^2 = f(r) dt^2 + \frac 1{f(r)} dr^2 + r^2 \big( d \theta^2 + \sin^2 \theta \ d\phi^2 \big) \qquad \text{where} \qquad f(r) = 1 - \frac{2m}r
\end{equation}
In this section, we will study the symplectic gauge fields corresponding to this metric, 
and then will study the geometry of the vector field tetrads arising from the gauge fields, and 
verify if it is self dual or not. For now, 
our first requirement will be to construct a new co-ordinate chart that will serve our purpose. 

\numberwithin{equation}{subsection}

\subsection{The Darboux chart}

The Darboux Theorem \cite{darboux} states that we can always locally eliminate dynamical gauge fields that 
fluctuate about the background vaccum condensate through a local co-ordinate transformation. 
In general relativity, the Equivalence Principle states that there always exists a diffeomorphism 
that equates a curved manifold locally to a flat manifold. This theorem applies for Riemannian 
manifolds.

Thus, the Darboux Theorem is the equivalence principle for Symplectic manifolds. It essentially 
states that the symplectic structure on a curved manifold can always be equated to the symplectic 
structure on a flat manifold via a diffeomorphism. It can be summed up by the mathematical 
statement below:
\begin{equation}
\label{dth} \exists \quad \frac{\partial y^\mu}{\partial \xi^a} \quad s.t. \quad \mathcal{F}_{\mu \nu} (x) \frac{\partial y^\mu}{\partial \xi^a} \frac{\partial y^\nu}{\partial \xi^b} = B_{ab}
\end{equation}
The question here is what kind of diffeomorphism will satisfy equation (\ref{dth}). The crudest answer 
we can give so far requires that we first write the perturbed symplectic structure as:
$$\mathcal{F}_{\mu \nu} (x) = B_{\mu \nu} + \lambda F_{\mu \nu} (x)$$
such that $\lambda$ sets the strength of the dynamical field perturbation to the symplectic structure. \\ \\
In the case of a given metric, we can compute the individual curvature components. Embedded 
within the curvature are the various $SU(2)_\pm$ gauge field components.
$$R_{ab} = \eta^{i(+)}_{ab} F^{i(+)} + \eta^{i(-)}_{ab} F^{i(-)} \qquad \Rightarrow \qquad F^{i(\pm)} = \frac14 \eta^{i(\pm)}_{ab} R_{ab}$$
The simplest way to eliminate local dynamical gauge fields upon switching to the Darboux 
co-ordinates, is to eliminate the individual $SU(2)_\pm$ gauge fields. This is necessarily true as we shall 
see below. It is known that in maximally symmetric spaces, we can have the curvature in the form:
$$R_{abcd} = g^{ij} (\vec{x}) \ \varepsilon_{iab} \ \varepsilon_{jcd}$$
In the case of self-dual curvature and fields, we can further elaborate it as:
$$R_{ab} = \alpha_{ij}^{(+)} (\vec{x}) \eta^{i(+)}_{ab} \eta^{j(+)}_{cd} + \alpha_{ij}^{(-)} (\vec{x}) \eta^{i(-)}_{ab} \eta^{j(-)}_{cd} \quad \Rightarrow \quad F^{i(\pm)} = \frac12 \alpha^{ij(\pm)} \eta^{j(\pm)}_{ab} e^a \wedge e^b$$
where all the $\alpha^{ij(\pm)} (\vec{x})$ tensor components are diagonal (ie. $\alpha^{ij(\pm)} (\vec{x}) = 0$ for $i \neq j$). This means 
that the dynamical gauge field strength affiliated with the metric as a linear combination of the 
individual components using the t'Hooft symbols as a basis.
$$F = c^{i(+)} F^{i(+)} + c^{i(-)} F^{i(-)}$$
$$\Rightarrow \qquad F_{ab} = c^{i(+)} \alpha^{ij(+)} (\vec{x}) \eta^{j(+)}_{ab} + c^{i(-)} \alpha^{ij(-)} (\vec{x}) \eta^{j(-)}_{ab}$$
Now, since these t'Hooft symbols never share the same non-zero matrix elements in the same 
positions, we can say that the $SU(2)_\pm$ gauge fields are linearly independent 2-forms. From linear 
algebra, we know that this implies that:
$$F = 0 \quad \longleftrightarrow \quad \alpha^{ij(\pm)} (\vec{x})  \quad \Rightarrow \quad F^{i(\pm)} = 0 \quad  \longleftrightarrow \quad R_{ab} = 0$$
This consequently eliminates the curvature as well, which describes the equivalence principle. Thus, 
if we can choose a local co-ordinate frame that locally eliminates the curvature, we will also have 
found the Darboux co-ordinates. We need local co-ordinates to obtain and analyse the gauge fields 
related to the metric. To do this, we could define a local co-ordinate system which preserves the 
volume element formed by the tetrads of (\ref{esmetric}).
\begin{equation} \label{volume} 
\begin{split}
\nu = \nu' &= \epsilon^1 \wedge \epsilon^2 \wedge \epsilon^3 \wedge \epsilon^4\\
\Rightarrow \quad e^1 \wedge e^2 \wedge e^3 \wedge e^4 &= d t \wedge \big( r^2 d r \big) \wedge \big( \sin \theta \ d \theta \big) \wedge d \phi
\end{split}
\end{equation}
These co-ordinates are known as the Darboux co-ordinates, the principle behind this design being 
to make the tetrads equivalent to the exact differentials of the local choice of co-ordinates.
\begin{equation}
\label{darboux} X^a = \big\{ \tau, \rho, x, y \big\} = \bigg\{ t, \frac{r^3}3, - \cos \theta, \phi \bigg\}
\end{equation}
The metric, in these co-ordinates are then written as:
\begin{equation}
\label{dmetric} ds^2 = \widetilde{f}(\rho) d \tau^2 + \frac 1{\widetilde{f}(\rho)} \frac{d \rho^2}{( 3 \rho )^{\frac43}} + ( 3 \rho )^{\frac23} \bigg\{ \frac{d x^2}{1 - x^2} + \big( 1 - x^2 \big) d y^2 \bigg\}, \qquad \qquad \widetilde{f}(\rho) = 1 - \dfrac{2m}{\big( 3 \rho \big)^{\frac13}}
\end{equation}
Thus, for the inverse tetrads we have:
\begin{align}
\bigg( \frac{\partial}{\partial s} \bigg)^2 &= \mathcal{E}_a \otimes \mathcal{E}_a = \lambda^{-2} V_a \otimes V_a \nonumber \\
\label{invdmetric} &= \widetilde{f}^{-1}(\rho) \bigg( \frac{\partial}{\partial \tau} \bigg)^2 + \widetilde{f}(\rho) (3 \rho)^{\frac43} \bigg( \frac{\partial}{\partial \rho} \bigg)^2 + \frac1{(3 \rho)^{\frac23}} \bigg\{ \big( 1 - x^2 \big) \bigg( \frac{\partial}{\partial x} \bigg)^2 + \frac1{\big( 1 - x^2 \big)} \bigg( \frac{\partial}{\partial y} \bigg)^2 \bigg\}
\end{align}
Looking at the metric (\ref{esmetric}) again, one can easily write the two matrices:
\begin{equation}
\label{vandinv}
\epsilon^a = \left({\begin{array}{cccc}
f^{\frac12}(r) & 0 & 0 & 0\\
0 & f^{-\frac12}(r) & 0 & 0\\
0 & 0 & r & 0\\
0 & 0 & 0 & r\sin \theta
\end{array} } \right) \qquad
\mathcal{E}_a = \left({\begin{array}{cccc}
f^{-\frac12}(r) & 0 & 0 & 0\\
0 & f^{\frac12}(r) & 0 & 0\\
0 & 0 & \frac1r & 0\\
0 & 0 & 0 & \frac1{r \sin \theta}
\end{array} } \right)
\end{equation}
Using the Darboux co-ordinates of (\ref{darboux}), we can define a symplectic form:
\begin{equation}
\label{sympform}
\omega = \epsilon^1 \wedge \epsilon^2 + \epsilon^3 \wedge \epsilon^4 \ = \ d \tau \wedge d \rho + d x \wedge d y \ = \ r^2 dt \wedge dr + \sin \theta \ d \theta \wedge d \phi 
\end{equation}
such that one can re-obtain the original volume form $\nu$
$$\nu = \frac12 \omega \wedge \omega = r^2 \sin \theta \ dt \wedge dr \wedge d \theta \wedge d \phi$$
that was shown in (\ref{volume}).

\subsubsection*{Complex Stereographic Projection - an alternate choice of coordinates}

Now it is understood that the polar co-ordinate system chosen here results in a multi-valuedness 
towards the poles that causes a breakdown of the one-to-one correspondence between the cartesian 
and polar variables, certifying a diffeomorphism, since the azimuthal angle $\phi$ is now arbitrary. 
$$\big( x, y, z \big) \longleftrightarrow \big( r, \theta, \phi \big) \hspace{1.5cm} \big( 0, 0, \pm r \big) \longleftrightarrow \big( r, 0, ? \big)$$
Thus, one needs to consider an alternate chart that preserves the correspondence. One such choice 
of local co-ordinates is the complex stereographic projection. There are two different charts for two different localities :
\begin{align}
\label{cster}
\mathbb{C} = U_+ = S^2 - \{x_\infty\} \quad : \quad \big( x, y, z \big) \longleftrightarrow \big( r, Z_+, \bar{Z}_+ \big) \quad \text{where} \quad Z_+ = \frac{x + iy}{r - z}\\
\bar{\mathbb{C}} = U_- = S^2 - \{x_0 \ \} \quad : \quad \big( x, y, z \big) \longleftrightarrow \big( r, Z_-, \bar{Z}_- \big) \quad \text{where} \quad Z_- = \frac{x - iy}{r + z}
\end{align}
where locality $\mathbb{C}$ describes the entire sphere except for the north pole, while $\bar{\mathbb{C}}$ describes the 
same sphere, only this time exempting the south pole, both with no arbitrary values in their localities:
$$U_- \ : \ \big( 0, 0, \ \ r \big) \longleftrightarrow \big( r, 0, 0 \big) $$
$$U_+ \ : \ \big( 0, 0, - r \big) \longleftrightarrow \big( r, 0, 0 \big) $$
The correspondence to the polar co-ordinates is given by:
\begin{equation}
\label{rule1} 
\begin{split}
Z_+ = \frac{e^{i \phi} \sin \theta}{1 - \cos \theta} &= e^{i \phi} \cot \frac{\theta}2, \qquad Z_- = \big( Z_+ \big)^{-1} \qquad d Z_+ = - \big( Z_- \big)^{-2} d Z_- \\
Z_+ \bar{Z}_- &= e^{2 i \phi} \qquad \big( \bar{Z}_+ \big)^{-1} Z_- = \tan \frac{\theta}2
\end{split}
\end{equation}
However, to preserve the volume element under this diffeomorphism we need to obtain the 
appropriate tetrad. This can be done by adjusting the wedge product:
\[ \begin{split}
- 2i \bigg( \frac{1 - \cos \theta}2 \bigg)^2 d Z_+ \wedge d \bar{Z}_+ &= \sin \theta \ d \theta \wedge d \phi \\
|Z_+|^2 = \frac{\sin^2 \theta}{(1 - \cos \theta)^2} = \frac{1 + \cos \theta}{1 - \cos \theta} &\quad \Rightarrow \quad 1 + |Z_+|^2 = \frac2{1 - \cos \theta}
\end{split} \]
\begin{equation}
\label{pveform} \therefore \quad \xi_+ = - 2i \frac{d Z_+ \wedge d \bar{Z}_+}{\big( 1 + |Z_+|^2 \big)^2 } = \sin \theta \ d \theta \wedge d \phi \hspace{0.75cm} \omega = * \xi_+ + \xi_+
\end{equation}
This 2-form holds the same (form invariant) expression in the other locality as well :
\begin{align}
\label{rule2}  d Z_+ \wedge d \bar{Z}_+ = | Z_- |^{-4} d Z_- \wedge d \bar{Z}_- &\hspace{2cm} 1 + | Z_+ |^2 = | Z_- |^{-2} \big( 1 + | Z_- |^2 \big)\\ 
\label{nveform} \therefore \quad \xi_- = - 2i &\frac{d Z_- \wedge d \bar{Z}_-}{ \big( 1 + | Z_- |^2 \big)^2 } \qquad \omega = *\xi_- + \xi_-
\end{align}
The respective volume element is given by:
$$\nu = \frac12 \omega \wedge \omega = - i \frac{ r^2 }{ \big( 1 + | Z_{\pm} |^2 \big)^2 } dt \wedge dr \wedge d Z_{\pm} \wedge d \bar{Z}_{\pm}$$
This 2-form's closure implies a potential field $A$, given by (\ref{sympform}), (\ref{pveform}) and (\ref{nveform}) as:
$$d \omega_{\pm} = 0 \quad \Rightarrow \quad \omega_{\pm} = d A_{\pm} = d \bigg( - \frac{r^3}3 dt + i \frac{Z_{\pm} d \bar{Z}_{\pm} - \bar{Z}_{\pm} d Z_{\pm}}{1 + |Z_{\pm}|^2} \bigg)$$
$$\Rightarrow \quad A_{\pm} = - \frac{r^3}3 dt + i \frac{Z_{\pm} d \bar{Z}_{\pm} - \bar{Z}_{\pm} d Z_{\pm}}{1 + |Z_{\pm}|^2} + d \varphi$$
Naturally, there is a chance of a constant or a first order exterior derivative seperating the two 
potential form representations. To describe the connection between $A_+$ and $A_-$ in the region 
$U_+ \cap U_-$, using (\ref{rule1}) and (\ref{rule2}) we have the following results:
\begin{equation}
d A_+ = d A_- \quad \Rightarrow \quad A_+ = A_- + d \varphi
\end{equation}
$$ Z_{+} d \bar{Z}_{+} - \bar{Z}_{+} d Z_{+} = - \frac{Z_{-} d \bar{Z}_{-} - \bar{Z}_{-} d Z_{-}}{|Z_-|^4}$$
\begin{equation}
A_+ + \frac{r^3}3 dt = - \frac1{|Z_-|^2} \bigg( A_- + \frac{r^3}3 dt \bigg) \quad \Rightarrow \quad A_+ - A_- = - i \frac{Z_{-} d \bar{Z}_{-} - \bar{Z}_{-} d Z_{-}}{|Z_-|^2}
\end{equation}
Now, we can say that for a complex number:
$$\frac{z \ d \bar{z} - \bar{z} \ dz}{|z|^2} = - 2 i \ d \big( \text{arg} (z) \big)$$
Thus, we can say that:
$$A_+ - A_- = - 2 \ d \big( \text{arg} (Z_-) \big) = 2 \ d \big( \text{arg} (Z_+) \big)$$
\begin{equation}
\boxed{A_+ = A_- + 2 \ d \big( \text{arg} (Z_+) \big)}
\end{equation}
Thus, as we can see that the two potentials for the two different localities, despite the same field 
strength form have a slight difference equivalent to the exterior derivative of the angular phase of 
the complex number. Now we proceed to obtain the symplectic gauge fields associated with the 
metric and study its salient properties.

\subsection{Symplectic Analysis}

Using the Darboux co-ordinates, we can obtain a symplectic gauge field set (recall eq.(\ref{sgf})):
$$C_a = B_{ab} X^b, \qquad  \theta^{ab} = \frac12 \eta^3_{ab} \quad \Rightarrow \quad B_{ab} = -2 \eta^3_{ab}$$
$$\text{where} \qquad \eta^3_{ab} = \left({\begin{array}{cccc}
0 & 1 & 0 & 0\\
-1 & 0 & 0 & 0\\
0 & 0 & 0 & 1\\
0 & 0 & -1 & 0
\end{array} } \right)$$
In matrix form the set of symplectic gauge fields are
$$C = -2 \left({\begin{array}{cccc}
0 & 1 & 0 & 0\\
-1 & 0 & 0 & 0\\
0 & 0 & 0 & 1\\
0 & 0 & -1 & 0
\end{array} } \right)
\left({\begin{array}{c}
\tau\\ \rho \\ x\\ y
\end{array} } \right) 
= -2 \left({\begin{array}{c}
\rho\\ -\tau\\ y\\ -x
\end{array} } \right) = -2 \left({\begin{array}{c}
\frac13 r^3\\ -t\\ \phi\\ \cos \theta
\end{array} } \right)$$
\begin{equation}
\label{gfields} \therefore \qquad C_1 = - \frac23 r^3, \qquad C_2 = 2 t, \qquad C_3 = -2 \phi \qquad C_4 = -2 \cos \theta
\end{equation}
We can now derive the vector fields corresponding to the symplectic gauge fields (\ref{gfields}) as the 
adjoint operation in the Poisson algebra and the result is shown in matrix form :
\begin{equation} 
\label{vecfld}
\begin{split}
V_a(f) &= \theta(C_a, f ) \qquad V_a^\mu = - \theta^{\mu \nu} \partial_\nu C_a \\ 
\therefore \quad V = \left({\begin{array}{cccc}
0 & 1 & 0 & 0\\
-1 & 0 & 0 & 0\\
0 & 0 & 0 & 1\\
0 & 0 & -1 & 0
\end{array} } \right)
&\left({\begin{array}{c}
\partial_t\\ \partial_r \\ \partial_\theta\\ \partial_\phi
\end{array} } \right)
\left({\begin{array}{cccc}
\frac13 r^3 & -t & \phi & \cos \theta
\end{array} } \right) = 
\left({\begin{array}{cccc}
r^2 & 0 & 0 & 0\\
0 & 1 & 0 & 0\\
0 & 0 & 1 & 0\\
0 & 0 & 0 & \sin \theta
\end{array} } \right)
\end{split}
\end{equation}
We have the formula to relate the vector field with the tetrads:
\begin{equation}
\label{lbda} V_a = \lambda E_a \qquad v^a = \lambda^{-1} e^a
\end{equation}
To determine the value of $\lambda$, we make use of the relation:
$$\lambda^2 = \ \text{det } V_a^\mu = r^2 \sin \theta \quad \Rightarrow \quad \lambda = r \sqrt{\sin \theta}$$
Now, the determinants of the volume preserving vector field array $V_a^\mu$ and that of the inverse vector 
field array, or corresponding tetrad array are given by:
$$\text{det} (V_a^\mu) =
\left|{\begin{array}{cccc}
r^2 & 0 & 0 & 0\\
0 & 1 & 0 & 0\\
0 & 0 & 1 & 0\\
0 & 0 & 0 & \sin \theta
\end{array} } \right|
= r^2 \sin \theta \qquad
\text{det} (V^a_\mu) =
\left|{\begin{array}{cccc}
\frac1{r^2} & 0 & 0 & 0\\
0 & 1 & 0 & 0\\
0 & 0 & 1 & 0\\
0 & 0 & 0 & \frac1{\sin \theta}
\end{array} } \right|
= \frac1{r^2 \sin \theta}$$
Knowing that $\lambda^2 = r^2 \sin \theta$ we can say that:
\begin{equation}
\text{det} (V^a_\mu) = \frac1{\lambda^2} \quad \Rightarrow \quad \lambda^2 = \frac1{\text{det} (V^a_\mu)} \quad \Rightarrow \quad v(x) = 1
\end{equation}
thus concluding that the inverse tetrad fields satisfy equation (5.145) of 
\cite{Yang:2013dec}.

\subsection{Bianchi identity for Symplectic gauge and Vector fields}

The Jacobi and Bianchi identities are well-studied in differential geometry. Both are derivatives of a basic identity defined by:
$$d^2 \omega^n = 0$$
where $\omega^n$ is an n-form. Having arisen from the 
same source, it is clear there is a connection between the two identities.

$$ \boxed{ \begin{split} 
\qquad \big\{ C_a , \big\{ C_b , C_c \big\}_\theta \big\}_\theta + \big\{ C_b , \big\{ &C_c , C_a \big\}_\theta \big\}_\theta + \big\{ C_c , \big\{ C_a , C_b \big\}_\theta \big\}_\theta = 0 \qquad \\
 &\big{\Updownarrow}\\
[V_a, [V_b, V_c]] + [V_b, [&V_c, V_a]] + [V_c, [V_a, V_b]] = 0 \\ 
\end{split} } $$

However, the above identities are valid only in regions where the metric is well defined. They break down in the presence of singularity as evident in electrodynamics where we find the Bianchi identity being invalid in the presence of static and dynamic
charge (current) distributions. 
\[ \begin{split}
 A = A_i \ dx^i \quad A_i = \{ \varphi, \vec{A} \} \qquad &F = dA \quad F_{ij} = \partial_i A_j - \partial_j A_i \equiv \{ \vec{E}, \vec{B} \} \\
\{ \rho, \vec{J} \} = \{0, \vec{0} \} : \quad dF = 0 \quad &\longrightarrow \quad \vec{\nabla} . \vec{E} = 0, \quad \vec{\nabla} . \vec{B} = 0\\
\{ \rho, \vec{J} \} \neq \{0, \vec{0} \} : \quad dF \neq 0 \quad &\longrightarrow \quad \vec{\nabla} . \vec{E} = \rho, \quad \vec{\nabla} . \vec{B} = 0
\end{split} \]
The Schwarzschild space with Lorentzian signature has an irremovable singularity at the origin, making the Bianchi identity invalid there. However, for the Euclidean signature metric, the singularity is removable \cite {espo} under Kruskal Szekeres co-ordinates which  means that for the Euclidean Schwarzschild instanton, the Bianchi identity is valid throughout the space. Also, remembering (\ref{bcurv}), we can conclude that:
$$\big[ V_a, \big[ V_b, V_c \big] \big] = 0 \qquad \Rightarrow \qquad {R_{abc}}^d = 0$$
showing that the local results are consistent with our emergent set-up.

\subsection{Seiberg Witten map and absence of self-duality}

Seiberg and Witten showed \cite{ncft-sw} that there are two equivalent descriptions - comutative 
and non-commutative of the low energy effective theory, depending on the regularization scheme 
or path integral prescription for the open string ending on a D-brane.

Since these two descriptions arise from the same open string theory depending on different 
regularizations, and the physics being independent of the regularization scheme, Seiberg and 
Witten argued that they should be equivalent. Thus there must be a spacetime field redefinition 
between ordinary and NC gauge fields, so called the Seiberg-Witten (SW) map.\\ \\
The relation for the NC field strength $\widehat{F}$ is given by \cite{Lee:2012rb}:
\begin{equation}
\big\{ C_a, C_b \big\}_\theta = -B_{ab} + \widehat{F}_{ab} \quad \Rightarrow \quad \widehat{F}_{ab} = B_{ab} + \big\{ C_a, C_b \big\}_\theta
\end{equation}
Using the $C$ matrix from (\ref{gfields}), we can write:
$$ \big\{ C_a, C_b \big\}_\theta = \left({\begin{array}{cccc}
0 & 2 r^2 & 0 & 0\\
-2 r^2 & 0 & 0 & 0\\
0 & 0 & 0 & 2 \sin \theta\\
0 & 0 & - 2 \sin \theta & 0
\end{array} } \right) $$
\begin{equation}
\label{Fhat} \therefore \qquad \widehat{F} = -2 \left({\begin{array}{cccc}
0 & 1 - r^2 & 0 & 0\\
- (1 - r^2) & 0 & 0 & 0\\
0 & 0 & 0 & 1 - \sin \theta\\
0 & 0 & - (1 - \sin \theta) & 0
\end{array} } \right)
\end{equation}
At this point, we recapitulate the Seiberg-Witten map between the field strengths of the two 
descriptions - commutative and non-commutative, given by the formula:
$$\widehat{F} = \big( 1 + F \theta \big)^{-1} F \quad \Rightarrow \quad F = \widehat{F} \big( 1 - \theta \widehat{F} \big)^{-1}$$
It is easy to see that the commutative gauge field strength $F_{\mu \nu}$
\begin{equation}
\label{fieldstrength}F = -2
\left({\begin{array}{cccc}
0 & \frac{1 - r^2}{r^2} & 0 & 0\\
-\frac{1 - r^2}{r^2} & 0 & 0 & 0\\
0 & 0 & 0 & \frac{1 - \sin \theta}{\sin \theta}\\
0 & 0 & -\frac{1 - \sin \theta}{\sin \theta} & 0
\end{array} } \right)
\end{equation}
 shows no self-duality at all, noncommutative or otherwise.

\subsection{The Seiberg-Witten field equation}

Now we consider the equation of motion of the gauge fields (\ref{fieldstrength}). We start by looking at the 
action corresponding to the gauge fields:
\begin{equation}
\label{swact} S = \frac1{4 g_{YM}} \int d^4 y \ \{ C_a, C_b \}^2
\end{equation}
$$\widehat{F} - B = (1  + F\theta).^{-1}\{F - B - F\} = - G^{-1} B$$
where we have chosen to substitute 
\begin{equation}
\label{swmet} G = 1  + F\theta = \left({\begin{array}{cccc}
r^{-2} & 0 & 0 & 0\\
0 & r^{-2} & 0 & 0\\
0 & 0 & (\sin \theta)^{-1} & 0\\
0 & 0 & 0 & (\sin \theta)^{-1}
\end{array} } \right)
\end{equation}
$$\therefore \qquad S = \int d^4 y \{ C_\mu , C_\nu \}_\theta^2 = \int d^4 x \sqrt{\text{Det}(G)} G^{\mu \lambda} G^{\nu \gamma} B_{\lambda \nu} B_{\gamma \mu}$$
The equation of motion can be obtained by minimising the variation of action (\ref{swact}):
$$\int d^4 y \big( \widehat{F} - B \big)^{\mu \nu} \big( \widehat{F} - B \big)_{\mu \nu} = \int d^4 x \sqrt{\text{Det}(G)} \big(G^{\mu \lambda} B_{\lambda \nu} \big) \big(G^{\nu \gamma} B_{\mu \gamma} \big)$$
$$\therefore  \qquad \int d^4 y \big( \widehat{F} - B \big)^{\mu \nu} \big( \widehat{F} - B \big)_{\mu \nu} = -\int d^4 x \sqrt{\text{Det}(G)} \text{Tr}(G^{-1} B G^{-1} B)$$
Noting that $A^{[\mu \nu]} = \frac 1 2 (A^{\mu \nu} - A^{\nu \mu})$, the commutative equation of motion is derived as:
\[\begin{split}
\delta S = 0 \qquad \Rightarrow& \qquad \delta \bigg[\int d^4 x \sqrt{\text{Det}(G)} \text{Tr}(G^{-1} B G^{-1} B)\bigg] = 0\\
\Rightarrow \qquad \int d^4 x \big[ \delta\big(\sqrt{\text{Det}(G)}\big) &\text{Tr}(G^{-1} B G^{-1} B) + \sqrt{\text{Det}(G)} . \delta \big\{\text{Tr}(G^{-1} B G^{-1} B)\big\}\big] = 0
\end{split}\]
In operator form, we write:
$$\delta\sqrt{\text{Det}(G)} = \frac 1 2 \sqrt{\text{Det}(G)} G^{-1} \delta(G) = \frac 1 2 \sqrt{\text{Det}(G)} \big(G^{-1}\big) \theta \ \delta F$$
\[\begin{split}
G^{-1}.G = \mathbb{I} \quad \Rightarrow \quad \delta\big((G^{-1})\big).G &= -G^{-1} . \delta (G) = -G^{-1} . \theta . \delta F\\
\Rightarrow \quad \delta\big((G^{-1})\big) &= -\big(\theta . G^{-1}\big) . \delta F . G^{-1}
\end{split}\]
Thus, the minimised action variation is:
\[\begin{split}
\therefore \qquad \int d^4 x \sqrt{\text{Det}(G)} \bigg[ \big(G^{-1}\big) \theta . \text{Tr}(G^{-1} B G^{-1} B)\delta F &+ 4 \text{Tr}( G^{-1} B \delta \big(G^{-1}\big) B) \bigg] = 0\\
\Rightarrow \qquad \int d^4 x \sqrt{\text{Det}(G)} \Big[ \big(\theta . G^{-1}\big) \text{Tr}(G^{-1}B G^{-1}B) &+ 4 ( G^{-1} B \big(\theta . G^{-1}\big) B G^{-1}) \Big]^{\mu \nu} \delta F_{\mu \nu} = 0
\end{split}\]
The variation of the gauge field $F$ and its application into the action variation are:
$$\delta F_{\mu \nu} = \delta(\partial_\mu A_\nu - \partial_\nu A_\mu) = \partial_\mu \delta A_\nu - \partial_\nu \delta A_\mu$$
\[\begin{split}
\therefore \qquad \int d^4 x \sqrt{\text{Det}(G)} \Big[ \big(\theta . G^{-1}\big) \text{Tr}(G^{-1}B G^{-1}B) &- 4 ( G^{-1} B \big(\theta . G^{-1}\big) B G^{-1}) \Big]^{[\mu \nu]} \partial_\mu \delta A_\nu = 0\\
\Rightarrow \qquad \partial_\mu \Big[ \sqrt{G} \big\{ \big( \theta G^{-1} \big)^{\mu \nu} \text{Tr}(G^{-1}B G^{-1}B) &- 4(\theta G^{-1}B G^{-1}B G^{-1})^{[\mu \nu]} \big\} \Big] = 0
\end{split}\]
Thus, the resulting equation of motion that is obtained for the first time here reads as:
\begin{equation}
\label{eom} \partial_\mu \Big[ \sqrt{G}\big\{(\theta G^{-1})^{\mu \nu} \text{Tr}(G^{-1}B G^{-1}B) - 4(\theta G^{-1}B G^{-1}B G^{-1})^{[\mu \nu]}\big\} \Big] = 0
\end{equation}
Substituting $G$ from (\ref{swmet}) into (\ref{eom}) above should give us the SW field equation for the Euclidean Schwarzschild metric which is a typical example of AF gravitational instaton.

\numberwithin{equation}{subsection}

\section{Geometric Analysis}

Now we proceed to analyze the various geometric and topological properties of the Euclidean 
Schwarzschild metric. This will involve obtaining the various topological invariants related to the 
metric. We will start by obtaining the curvature components of the metric. 

\subsection{Curvature analysis}

We can extract the complete set of tetrads for the metric (\ref{esmetric}) as:
\begin{equation}
\begin{split}
\label{eset} e^1 &= \sqrt{1 - \frac{2m}r} dt \hspace{2cm} e^2 = \frac 1{\sqrt{1 - \frac{2m}r}} dr\\
e^3 &= r d\theta \hspace{3.5cm} e^4 = r \sin \theta \ d\varphi
\end{split}
\end{equation}
Starting with (\ref{eset}) and using Cartan's 1st torsion-free structure equation, we have:
\[ \begin{split}
{\omega^1}_2 = -{\omega^2}_1 &= \frac{m}{r^2}dt \hspace{3.3cm} {\omega^4}_3 = -{\omega^3}_4 = \cos \theta \ d\varphi \\
{\omega^3}_2 = -{\omega^2}_3 &= \sqrt{1 - \frac{2m}r} d\theta \hspace{2cm} {\omega^4}_2 = -{\omega^2}_4 = \sqrt{1 - \frac{2m}r} \sin \theta \ d\varphi
\end{split} \]
The overall $\omega$ (spin-connection) matrix is given by:
\begin{equation}
\label{spin} {\omega^i}_j = \frac1r
\left({\begin{array}{cccc}
0 & a \ e^1 & 0 & 0\\
- a \ e^1 & 0 & - b \ e^3 & - b \ e^4\\
0 &  b \ e^3 & 0 & - c \ e^4\\
0 &  b \ e^4 & c \ e^4 & 0
\end{array} } \right)
\qquad \text{where} \qquad
\begin{cases}
a = \frac{m}{r \sqrt{f(r)}} \\ b = \sqrt{f(r)} \\ c = \cot \theta
\end{cases}
\end{equation}
For the curvature components, we use the 2nd structure equation:
\begin{equation}
\label{curv} {R^i}_j = d {\omega^i}_j + {\omega^i}_k \wedge {\omega^k}_j
\end{equation}
Combining (\ref{spin}) and (\ref{curv}) gives the following non-vanishing curvature components:
\begin{equation} \label{riemcom}
\begin{split}
{R^1}_{212} = -{R^1}_{221} = -{R^2}_{112} = {R^2}_{121} = \frac{2m}{r^3} \\
{R^1}_{313} = -{R^1}_{331} = -{R^3}_{113} = {R^3}_{131} = -\frac{m}{r^3} \\
{R^1}_{414} = -{R^1}_{441} = -{R^4}_{114} = {R^4}_{141} = -\frac{m}{r^3} \\
{R^2}_{323} = -{R^2}_{332} = -{R^3}_{223} = {R^3}_{232} = - \frac{m}{r^3} \\
{R^2}_{424} = -{R^2}_{442} = -{R^4}_{224} = {R^4}_{242} = - \frac{m}{r^3} \\
{R^3}_{434} = -{R^3}_{443} = -{R^4}_{334} = {R^4}_{343} = \frac{2m}{r^3}
\end{split}
\end{equation}
In a compact form, the $R_{ab}$ matrix can be written as:
\begin{equation}
\label{rmatrix} R_{ab} = 
\frac{m}{r^3} \left({\begin{array}{cccc}
0 & 2 x & - y & - z\\
- 2 x & 0 & - \bar{z} & \bar{y}\\
y & \bar{z} & 0 & 2 \bar{x}\\
z & - \bar{y} & - 2 \bar{x} & 0
\end{array} } \right)
\end{equation}
where we use the representation:
\[ \begin{split}
x = e^1 \wedge e^2 \qquad &y = e^1 \wedge e^3 \qquad z = e^1 \wedge e^4 \\
\bar{x} = e^3 \wedge e^4 \qquad &\bar{y} = e^4 \wedge e^2 \qquad \bar{z} = e^2 \wedge e^3 \\
\text{where} \qquad x &\wedge \bar{x} = y \wedge \bar{y} = z \wedge \bar{z} = \nu
\end{split} \]
where $\nu$ is the volume form. Clearly, we can see that $R_{ab}$ matrix of (\ref{rmatrix}) is not self dual since each of its 
components are made of only one 2-form term, making it impossible to exhibit self-duality.
$$*R_{ab} = \frac12 \frac{{\varepsilon_{ab}}^{cd}}{\sqrt{g}} R_{cd} \neq R_{ab}$$
Now with the Riemann tensor components from (\ref{riemcom}), we can compute the Ricci tensor and scalar
$$R_{ij} = \eta^{kl} R_{ikjl} = \eta_{im} \eta^{kl} {R^m}_{kjl} \qquad \Rightarrow \qquad R_{11} = R_{22} = R_{33} = R_{44} = 0$$
\begin{equation}
\label{scalar} R = \eta^{ij} R_{ij} \qquad \Rightarrow \qquad R = 0
\end{equation}
So the Euclidean Schwarzschild solution classified in the literature as AF gravitational instanton
does not exhibit self-duality although it is a Ricci-flat manifold. Since the spin connections in eq. (\ref{spin}) are 
neither self-dual or anti-self dual, we can proceed to construct both type of SU(2) gauge fields
and the field strengths using respectively the spin connections (\ref{spin}) and curvature 
components (\ref{rmatrix}) using the following formula:
\begin{equation}
\label{gf} A^{(\pm)i} = \frac14 \eta^{(\pm)i}_{\mu \nu} \omega_{\mu \nu} \qquad \qquad F^{(\pm)i} = \frac14 \eta^{(\pm)i}_{\mu \nu} R_{\mu \nu}
\end{equation}
By construction the field strengths should be either self-dual (for the + sign) or anti-self dual (for 
the - sign). According to a general result (3.41) found in \cite{ohparkhsyang}, the SU(2) gauge field 
(\ref{gf}) automatically satisfy the self duality equation and hence these solution describes an SU(2) Yang-Mills (anti) instanton on the space (\ref{esmetric}). \\ \\
Thus, we have the following description for the  $SU(2)_+$ instanton and $SU(2)_-$ anti-instanton gauge fields respectively listed as :
\begin{equation}
\label{gpot}
\begin{split}
A^{(+)1} &=  - \frac1{2r} b \ e^3 \hspace{4cm}
A^{(-)1} = - \frac b{2r} \ e^3\\
A^{(+)2} &= - \frac1{2r} b \ e^4 \hspace{4cm}
A^{(-)2} = - \frac b{2r} \ e^4\\
A^{(+)3} &= \frac1{2r} \big( a \ e^1 - c \ e^4 \big) \hspace{2.7cm}
A^{(-)3} = \frac1{2r} \big( a \ e^1 + c \ e^4 \big)
\end{split}
\end{equation}
\begin{equation}
\label{gfield}
\begin{split}
F^{(+)1} &= - \frac{m}{2r^3} \big( z + \bar{z} \big) \hspace{4cm}
F^{(-)1} = \frac{m}{2r^3} \big( z - \bar{z} \big)\\
F^{(+)2} &= \frac{m}{2r^3} \big( y + \bar{y} \big) \hspace{4.3cm}
F^{(-)2} = \frac{m}{2r^3} \big( y - \bar{y} \big)\\
F^{(+)3} &= \frac{m}{r^3} \big( x + \bar{x} \big) \hspace{4.5cm}
F^{(-)3} = \frac{m}{r^3} \big( x - \bar{x} \big)
\end{split}
\end{equation}
Remembering that the curvature components are given by (\ref{curv}), we can write:
$${R^a}_b = \frac12 {R^a}_{b \mu \nu} \ dx^\mu \wedge dx^\nu \quad \Rightarrow \quad {R^a}_{bcd} = \iota_{E_d} \iota_{E_c} {R^a}_b$$
$$\iota_{E_d} \iota_{E_c} \big( d {\omega^a}_b + {\omega^a}_m \wedge {\omega^m}_b \big) = \big\{ \partial_c ( {{\omega_d}^a}_b ) - {{\omega_c}^m}_d \ {{\omega_m}^a}_b + {{\omega_c}^a}_m \ {{\omega_d}^m}_b \big\}$$
$$\therefore \quad {R^a}_{bcd} = \big\{ \nabla_c ( {{\omega_d}^a}_b ) - {{\omega_c}^m}_d \ {{\omega_m}^a}_b + {{\omega_c}^a}_m \ {{\omega_d}^m}_b \big\}$$
Thus we get the Ricci tensor to be
\[ R_{ac} = \big\{ \underbrace{ \nabla_c ( f_{bab} ) }_0 - \omega_{cmb} \ \omega_{mab} - \omega_{cma} \ f_{bmb} \big\} \]
Finally the Ricci scalar can be written as
$$\quad R = - \big\{ \omega_{amb} \ \omega_{mab} + f_{ama} \ f_{bmb} \big\}$$
Now, since the Ricci scalar vanishes in our case (see eqn. (\ref{scalar})), we have:
\begin{equation}
\label{ricci} R = 0 \quad \Rightarrow \quad (f_{aba})^2 = \omega_{abc} \ \omega_{cab}
\end{equation}
The tetrads and the vector fields in (\ref{vecfld})-(\ref{lbda}) exhibit the structure equations:
\begin{equation}
\label{structure}
\big[ E_a, E_b \big] = - {f_{ab}}^c E_c \qquad \big[ V_a, V_b \big] = - {g_{ab}}^c V_c
\end{equation}
If the vector fields $\{E_a\}$ and $\{V_a\}$ are related by (\ref{lbda}), then we can suppose that:
$$d V_b = d \big( \lambda E_b \big) = d \lambda \wedge E_b - \lambda {\omega^c}_b E_c = d ( \log{\lambda} ) \wedge V_b - {\omega^c}_b V_c$$
$$\iota_{V_a} dV_b = V_a V_b = V_a ( \log{\lambda} ) V_b - \lambda \ {{\omega_a}^c}_b  V_c$$
$$\therefore \quad \big[ V_a, V_b \big] = V_a ( \log{\lambda} ) V_b - V_b ( \log{\lambda} ) V_a - \lambda \big( {{\omega_a}^c}_b - {{\omega_b}^c}_a \big)  V_c$$
$$\Rightarrow \quad - {g_{ab}}^c V_c = \frac12 \big( {g_{ma}}^m V_b - {g_{mb}}^m V_a \big) - \lambda \ {f_{ab}}^c  V_c $$
So we can write the structure constants in terms of the metric
\begin{equation}
\label{struct} {f_{ab}}^c = \frac1{\lambda} \bigg\{ {g_{ab}}^c + \frac12 \bigg({g_{ma}}^m \delta_b^c - {g_{mb}}^m \delta_a^c \bigg) \bigg\} \quad \Rightarrow \quad {f_{ab}}^a = \frac1{\lambda} {g_{ab}}^a
\end{equation}
We also note the relation between spin connection and structure constant:
\begin{equation}
\label{spstr} \omega_{abc} = \frac12 \Big( f_{abc} - f_{bca} + f_{cab}  \Big)
\end{equation}
Finally, an important identity here is:
\begin{align}
\label{sdc1}
\rho^b = {g_a}^{ba} \quad \quad &\Psi^d = \frac12 \varepsilon^{abcd} g_{abc}\\
\label{sdc2}
\rho_b \rho^b = \Psi_d \Psi^d \quad & \Rightarrow \quad \rho^a = \pm \Psi^a
\end{align}
With a little effort, it can be shown (in any $2n$-dimensions) \cite{hsy-jpcs12, hsyangjhep09} that the right-hand side of
the Bianchi identity for vector fields is precisely equivalent to the first Bianchi identity of Riemann curvature tensors, i.e.,
\begin{equation} \label{1-bianchi}
 [V_a, [V_b, V_c]] + \mathrm{cyclic} = 0  \qquad \Leftrightarrow
 \qquad R_{[abc]d} = 0,
\end{equation}
where $[abc]$ denotes the cyclic permutation of indices. The equation (\ref{1-bianchi}) leads to a cryptic result 
for Ricci tensors \cite{hsy-jpcs12,hsyangjhep09}
\begin{equation} \label{emergent-einstein}
  R_{ab} = - \frac{1}{\lambda^2} \Big[ g^{(+)i}_d g^{(-)j}_d
  \Big(\eta^i_{ac} \overline{\eta}^{j}_{bc}
  + \eta^i_{bc} \overline{\eta}^{j}_{ac} \Big) - g^{(+)i}_c
  g^{(-)j}_d \Big(\eta^i_{ac} \overline{\eta}^{j}_{bd}
  + \eta^i_{bc} \overline{\eta}^{j}_{ad} \Big) \Big]
\end{equation}
where $\eta^i_{ab}$ and $\overline{\eta}^{i}_{ab}$ are self-dual and
anti-self-dual 't Hooft symbols. To get the result
(\ref{emergent-einstein}), we have to define the canonical
decomposition of the structure equation (\ref{structure}) like
\begin{equation}\label{def-g} 
g_{abc} = g^{(+)i}_c \eta^i_{ab} + g^{(-)i}_c
\overline{\eta}^{i}_{ab}.
\end{equation}
A notable point is that the right-hand side of
(\ref{emergent-einstein}) consists of purely interaction terms
between self-dual and anti-self-dual parts in (\ref{def-g}) which is
the feature withheld by matter fields only \cite{ohhsyang}. A
gravitational instanton which is a Ricci-flat, K\"ahler manifold can
be understood as either $g^{(-)i}_c = 0$ (self-dual) or $g^{(+)i}_c = 0$ 
(anti-self-dual) in terms of (\ref{def-g}) and so $R_{ab} = 0$
in (\ref{emergent-einstein}). Hence, the result
(\ref{emergent-einstein}) is consistent with the Ricci-flatness of
gravitational instantons. However (\ref{emergent-einstein}) also has a nontrivial trace
contribution, i.e., a nonzero Ricci scalar, due to the second part
which does not exist in Einstein gravity. The content of the energy-momentum tensor defined by the
right-hand side of the Bianchi identity for vector fields becomes manifest by
decomposing it into two parts, denoted by $8\pi G T_{ab}^{(M)}$ and
$8\pi G T_{ab}^{(L)}$, respectively \cite{hsy-jpcs12,hsyangjhep09}:
\begin{eqnarray} \label{emt-max}
8\pi G T_{ab}^{(M)} &=& - \frac{1}{\lambda^2} \Big( g_{acd} g_{bcd}
- \frac{1}{4} \delta_{ab} g_{cde} g_{cde} \Big), \\
\label{emt-liouville}
8\pi G T_{ab}^{(L)} &=&  \frac{1}{2 \lambda^2} \Big( \rho_a
\rho_b - \Psi_a \Psi_b - \frac{1}{2}\delta_{ab} \big(\rho_c^2 -
\Psi_c^2 \big) \Big),
\end{eqnarray}
\begin{equation}\label{rho-psi}
  \text{where} \hspace{2cm}  \rho_a \equiv g_{bab}, \qquad \Psi_a \equiv - \frac{1}{2}
    \varepsilon^{abcd} g_{bcd}.
\end{equation}
The first energy-momentum tensor (\ref{emt-max}) is traceless, i.e.
$8\pi G T_{aa}^{(M)} = 0$, which is a consequence of the identity
$\eta^i_{ab} \overline{\eta}^{j}_{ab} = 0$ when applied to the first
part of (\ref{emergent-einstein}).  The Ricci scalar $R
\equiv R_{aa}$ can be calculated by (\ref{emt-liouville}) to
yield
\begin{equation}\label{r-scalar}
R = \frac{1}{2 \lambda^2} \Big( \rho_a^2 - \Psi_a^2 \Big).
\end{equation}
The equation (\ref{r-scalar}) immediately leads to the conclusion
that a four-manifold emergent from pure symplectic gauge fields
(without source terms) can have a vanishing Ricci scalar if and only
if (see eqn. (\ref{sdc1}) and (\ref{sdc2}) and its derivation)
\begin{equation}\label{scalar-dual}
    \rho_a = \pm \Psi_a
\end{equation}
that is similar to the self-duality equation. When the relation (\ref{scalar-dual}) is obeyed, the second energy-
momentum tensor $8\pi G T_{ab}^{(L)}$ (\ref{emt-liouville}) identically vanishes which confirms that the space of 
a Euclidean Schwarzschild solution is complete vacuum with no matter present.

\subsection{Topological Invariants}

\numberwithin{equation}{subsection}

In gravity topology can play a role at various levels. At the macroscopic level one may consider
multiplying corrected universes and wormholes, whilst at the microscopic Planck scale spacetime
topology may subject to quantum fluctuations; in analogy with others QFTs like sigma models and 
Yang-Mills theories, it is expected that the quantum tunneling process between different 
topologies are dominated by finite-action solutions of Euclidean gravity, the gravitational 
instantons.

One way to characterize topologically non-trivial solutions of the gravitational field equations is by
the value of topologically invariant integral over certain polynomials of the curvature tensor. In four
dimensions there are essentially two independent topological invariants the Euler Charcteristics 
and the Hirzebruch signature \cite{gibhawk4}. Every manifold with an associated metric has topological invariants 
that characterize it, implying geometric similarities between manifolds sharing the same invariant. 
Here, we will calculate two topological invariants of the Euclidean Schwarzschild instanton.

\subsubsection{Euler characteristic}

We can use the Riemann tensor components to compute the Euler characteristic given by:
\begin{equation}
\label{euler}
\chi(M) = \frac 1{32\pi^2} \int_M \varepsilon^{abcd} R_{ab} \wedge R_{cd} + \frac1{16 \pi^2}\int_{\partial M} \varepsilon^{abcd} \bigg( \theta_{ab} \wedge R_{cd} - \frac23 \theta_{ab} \wedge \theta_{cp} \wedge \theta_{pd} \bigg)
\end{equation}
where $\theta_{AB}$ is the second fundamental form of the boundary $\partial M$. It is defined by
\begin{equation} \label{2-f-form}
\theta_{AB} = \omega_{AB} - \omega_{0AB},
\end{equation}
where $\omega_{AB}$ are the actual connection 1-forms and $\omega_{0AB}$ are the connection 1-forms if the metric 
were locally a product form near the boundary \cite{egh-report}. The connection 1-form $\omega_{0AB}$ will have only 
tangential components on $\partial M$ and so the second fundamental form $\theta_{AB}$ will have only normal 
components on $\partial M$. The bulk part of the Euler characteristic is given by:
\begin{equation}
\label{ebulk}
\chi_{bulk} = \frac 1{32\pi^2} \int_M \varepsilon^{abcd} R_{ab} \wedge R_{cd}
\end{equation}
To compute the expression in (\ref{ebulk}), we only need to consider 6 combinations, where one half is 
equivalent to the other half. These combinations are given as:
\begin{equation}
\label{RwR}
\begin{split}
R_{12} \wedge R_{34} &= R_{34} \wedge R_{12}\\
R_{13} \wedge R_{24} &= R_{24} \wedge R_{13}\\
R_{14} \wedge R_{23} &= R_{23} \wedge R_{14}
\end{split}
\end{equation}
Since each permutation of 2 index pairs yields 2 combinations, and as shown in (\ref{RwR}), equivalent 
pairs of combinations exist, we can say that (\ref{ebulk}) reduces to:
\begin{equation}
\label{cbulk} \chi_{bulk} = \frac 1{4\pi^2} \int_M \Big( \varepsilon^{1234} R_{12} \wedge R_{34} + \varepsilon^{1324} R_{13} \wedge R_{24} + \varepsilon^{1423} R_{14} \wedge R_{23} \Big)
\end{equation}
We can use the Bianchi identity for curvature tensor to show that:
$$R_{ab} \wedge R_{cd} = d \omega_{ab} \wedge R_{cd} + \omega_{ap} \wedge {\omega^p}_b \wedge R_{cd}$$
$$d \omega_{ab} \wedge R_{cd} = d \big( \omega_{ab} \wedge R_{cd} \big)$$
$$\omega_{am} \wedge {\omega^m}_b \wedge R_{cd} = \omega_{ap} \wedge {\omega^p}_b \wedge d \omega_{cd} + \omega_{ap} \wedge {\omega^p}_b \wedge \omega_{cq} \wedge {\omega^q}_d$$
\begin{equation}
\label{oor}
\therefore \qquad \omega_{ap} \wedge {\omega^p}_b \wedge R_{cd} = d \big( \omega_{ap} \wedge {\omega^p}_b \wedge \omega_{cd} \big) + \omega_{ap} \wedge {\omega^p}_b \wedge \omega_{cq} \wedge {\omega^q}_d
\end{equation}
\begin{align} 
\int_M R_{ab} \wedge R_{cd} = \int_M &d \big( \omega_{ab} \wedge R_{cd} + \omega_{am} \wedge {\omega^m}_b \wedge \omega_{cd} \big) + \int_M \omega_{am} \wedge {\omega^m}_b \wedge \omega_{cn} \wedge {\omega^n}_d \nonumber\\
\label{expan} = \int_{\partial M} \big( \omega_{ab} &\wedge R_{cd} + \omega_{am} \wedge {\omega^m}_b \wedge \omega_{cd} \big) + \int_M \omega_{am} \wedge {\omega^m}_b \wedge \omega_{cn} \wedge {\omega^n}_d
\end{align}
We can see that for the 2nd term in (\ref{oor}) and for the 3rd term in (\ref{expan}) that:
$$\varepsilon^{abcd} \omega_{ap} \wedge {\omega^p}_b = \varepsilon^{abcd} \big( \omega_{ac} \wedge {\omega^c}_b + \omega_{ad} \wedge {\omega^d}_b \big)$$
\begin{equation}
\label{z1} \therefore \quad \varepsilon^{abcd} \omega_{ap} \wedge {\omega^p}_b \wedge \omega_{cq} \wedge {\omega^q}_d = 0
\end{equation}
Using (\ref{z1}) we can see that (\ref{expan}) becomes:
\begin{equation}
\label{result} \int_M R_{ab} \wedge R_{cd} = \int_{\partial M} \big( \omega_{ab} \wedge R_{cd} + \omega_{am} \wedge {\omega^m}_b \wedge \omega_{cd} \big)
\end{equation}
For the 2nd term, we refer to (\ref{spin}) to point out that besides the 2nd row and column, all other 
rows and columns have only 2 non-zero elements (the first one has only one). ie.:
$$\sum_m \varepsilon^{abcd} \omega_{ap} \wedge {\omega^p}_b \wedge \omega_{cd} = \varepsilon^{abcd} \Big( \omega_{ac} \wedge {\omega^c}_b \wedge \omega_{cd} + \omega_{ad} \wedge {\omega^d}_b \wedge \omega_{cd} \Big) = 0; \quad \forall \ \  c, d \neq 2$$
Thus, the different non-vanishing components of (\ref{result}) are:
\begin{align}  
\label{pair1} \int_M R_{12} \wedge R_{34} &= \int_{\partial M} \omega_{12} \wedge R_{34} = \int_{\partial M} \frac{2m^2}{r^3} dt \wedge d \theta \wedge \sin \theta d \phi \\
\label{pair2} \int_M R_{13} \wedge R_{24} &= - \int_{\partial M} \omega_{12} \wedge {\omega^2}_3 \wedge \omega_{24} = - \int_{\partial M} \frac{m}{r^2} \bigg( 1 - \frac{2m}r \bigg) dt \wedge d \theta \wedge \sin \theta d \phi \xrightarrow{r = 2m} 0 \\
\label{pair3} \int_M R_{14} \wedge R_{23} &= \int_{\partial M} \omega_{12} \wedge {\omega^2}_4 \wedge \omega_{23} = - \int_{\partial M} \frac{m}{r^2} \bigg( 1 - \frac{2m}r \bigg) dt \wedge d \theta \wedge \sin \theta d \phi \xrightarrow{r = 2m} 0
\end{align}
Applying (\ref{result}), (\ref{pair1}), (\ref{pair2}) and (\ref{pair3}) to (\ref{cbulk}) gives us:
\begin{equation}
\label{rbulk} \chi_{bulk} = \frac 1{4 \pi^2} \int_{\partial M} \omega_{12} \wedge R_{34} = \frac 1{4 \pi^2} \frac{2m^2}{r_h^3} \int_0^\beta dt \wedge \int_0^\pi \sin \theta d \theta \wedge \int_0^{2 \pi} d \phi = \frac{2m^2}{\pi r_h^3} \beta
\end{equation}
Here, we compactify the imaginary time, such that it lies within the range: $0 \leq t \leq \beta$ (generalization 
of the condition of the removal of conical singularity for our class of metrics). The upper limit $\beta$ 
(realized as inverse temperature for the black hole) is given by:
$$\kappa \beta = 2 \pi \qquad \text{where} \quad \kappa = \frac12 \frac{\partial_r g_{tt}}{\sqrt{g_{tt} g_{rr}}} \bigg\vert_{r = r_h} = \frac12 \big( \partial_r f(r) \big)_{r = r_h} = \frac m{r_h^2}$$
\begin{equation}
\label{tlim} \therefore \quad \beta = \frac{2 \pi}{\kappa} = \frac{2 \pi r_h^2}{m}
\end{equation}
Thus, for  Schwarzschild, $r_h = 2m$ and applying (\ref{tlim}) in (\ref{rbulk}) the bulk part is:
$$\chi_{bulk} = \frac{4m}{r_h} = 2$$
The boundary integral term of the Euler characteristics is given by:
$$\chi_{boundary} = \frac1{16 \pi^2}\int_{\partial M} \varepsilon^{abcd} \bigg( \theta_{ab} \wedge R_{cd} - \frac23 \theta_{ab} \wedge \theta_{cp} \wedge \theta_{pd} \bigg)$$
Recall that, the 1-form $\theta_{ab}$ is given by:
$$\theta_{ab} = \omega_{ab} - \omega_{0ab}, \hspace{2cm} \text{where } \qquad \omega_{0ab} = \big( \omega_{ab} \big)_{r = \infty}$$
Only the component along the normal to the surface is to be treated differently ie.:
$$\theta_{12} = \omega_{12}$$
The ${\theta^a}_b$ matrix is given by:
$${\theta^a}_b = 
\left({\begin{array}{cccc}
0 & \frac{m}{r^2}dt & 0 & 0\\
-\frac{m}{r^2}dt & 0 & \bigg(1 - \sqrt{1 - \frac{2m}r} \bigg) d\theta & \bigg(1 - \sqrt{1 - \frac{2m}r} \bigg) \sin \theta \ d\varphi\\
0 & - \bigg(1 - \sqrt{1 - \frac{2m}r} \bigg) d\theta & 0 & 0\\
0 & - \bigg(1 - \sqrt{1 - \frac{2m}r} \bigg) \sin \theta \ d\varphi & 0 & 0
\end{array} } \right)$$
In this case, since $\partial M \quad \Rightarrow \quad r = \infty$, when $\theta_{12}$ vanishes as $r \rightarrow \infty$. Thus, we can effectively say, 
$\theta_{ab} = 0$ which corresponds to setting $\chi_{boundary} = 0$ so that we can write:
\begin{equation}
\chi(M) = \chi_{bulk} + \chi_{boundary} = 2 + 0 = 2
\end{equation}
which is the value of Euler characteristic for Euclidean Schwarzschild metric. (see also \cite
{Liberati:1995jj} for a similar computation which was reported there for the first time.) \\ \\
Recalling how the Schwarzschild metric is a sum of an $SU(2)_{L}$  instanton and $SU(2)_{R}$  anti-instanton 
resulting from the $SU(2)_+$ and $SU(2)_-$gauge fields (described in the appendix (\ref{gf1}), (\ref{gf2}) and 
(\ref{gf3})), we can further calculate the Euler characteristics using:
$$\eta^{(\pm)i}_{\mu \nu} \eta^{(\pm)i}_{\lambda \gamma} = \delta_{\mu \lambda} \delta_{\nu \gamma} - \delta_{\mu \gamma} \delta_{\nu \lambda} \pm \varepsilon_{\mu \nu \lambda \gamma} \qquad \Rightarrow \qquad \varepsilon_{\mu \nu \lambda \gamma} = \frac12 \Big( \eta^{(+)i}_{\mu \nu} \eta^{(+)i}_{\lambda \gamma} - \eta^{(-)i}_{\mu \nu} \eta^{(-)i}_{\lambda \gamma} \Big)$$
Thus (\ref{ebulk}) reduces to
$$\frac1{32 \pi^2} \int_M \varepsilon^{abcd} R_{ab} \wedge R_{cd} = \frac1{4 \pi^2} \int_M \Big( F^{(+)i} \wedge F^{(+)i} - F^{(-)i} \wedge F^{(-)i} \Big)$$
It is straightforward to express the topological invariant in terms of $SU(2)$ gauge fields.
\begin{equation}
\label{gfvint} \therefore \quad \chi_{bulk} = \frac1{4 \pi^2} \int_M \Big( F^{(+)i} \wedge F^{(+)i} - F^{(-)i} \wedge F^{(-)i} \Big)
\end{equation}
We could now follow the same process as before invoking Stoke's theorem and convert (\ref{gfvint}) into 
a boundary integral using (\ref{result}) to obtain:
\[\begin{split}
\frac{\varepsilon^{abcd}}{32 \pi^2} \int_M R_{ab} \wedge R_{cd} &= \frac{\varepsilon^{abcd}}{32 \pi^2} \int_{\partial M} \big( \omega_{ab} \wedge R_{cd} + \omega_{am} \wedge {\omega^m}_b \wedge \omega_{cd} \big)\\
&= \frac1{4 \pi^2} \int_{\partial M} \Big( A^{(+)i} \wedge F^{(+)i} - A^{(-)i} \wedge F^{(-)i} \Big) + \frac{\varepsilon^{abcd}}{32 \pi^2} \int_{\partial M} \omega_{am} \wedge {\omega^m}_b \wedge \omega_{cd}
\end{split}\]
Seeing how the 2nd integrand vanishes for most combinations, and otherwise vanishes on the boundary 
itself, we can focus on the 1st integrand alone.
\begin{equation}
\label{gfbint} \chi_{bulk} = \frac1{4 \pi^2} \int_{\partial M} \Big( A^{(+)i} \wedge F^{(+)i} - A^{(-)i} \wedge F^{(-)i} \Big) = \chi^+_{bulk} + \chi^-_{bulk}
\end{equation}
Thus, for (\ref{gfbint}) we can compute the Euler character bulk values using (\ref{gpot}) and (\ref{gfield}) as:
\[ \begin{split}
\chi^+_{bulk} &= \frac{m^2}{2 r_h^3 \pi} \beta + \frac{m}{4 r_h^2 \pi} \beta = \bigg( \frac1{16 m \pi} + \frac1{16 m \pi} \bigg) \beta = 1 \\ 
\chi^-_{bulk} &= - \frac1{4 \pi^2} \int_{\partial M} - \frac m{4 r^4} \big( b e^3 \wedge z - b e^4 \wedge y + 2 a e^1 \wedge \bar{x} - 2 c e^4 \wedge x \big) = 1
\end{split} \]
For verification, we evaluate the contributions according to (\ref{gfvint}) using (\ref{gfield}) to get:
\[ \begin{split}
\chi^+_{bulk} &= \frac1{4 \pi^2} \int_M F^{(+)i} \wedge F^{(+)i} = \frac{m^2}{r_h^3 \pi} \beta = \frac{2 m}{r_h} = 1 \\
\chi^-_{bulk} &= - \frac1{4 \pi^2} \int_M F^{(-)i} \wedge F^{(-)i} = - \frac1{4 \pi^2} \int_M - \bigg( \frac{m^2}{2 r^6} + \frac{m^2}{2 r^6} + \frac{2 m^2}{r^6} \bigg) \nu = 1
\end{split} \]
Thus, we can clearly see that the overall bulk value of the Euler characteristic is the sum of the two 
individual values due to $SU(2)_+$ and $SU(2)_-$ gauge fields, giving:
\begin{equation}
\chi_{bulk} = \chi^+_{bulk} + \chi^-_{bulk} = 1 + 1 = 2
\end{equation}
This also shows both gauge fields contributing eqally to the overall Euler invariant.

\subsubsection{Hirzebruch signature}

Now we turn our attention to the other topological invariant, the Hirzebruch signature, given by
\begin{equation}
\label{hsig} \tau(M) = - \frac1{24 \pi^2} \bigg( \int_M \text{Tr} \ R \wedge R + \int_{\partial M} \text{Tr} \ \theta \wedge R + \eta_S ( \partial M ) \bigg)
\end{equation}
The bulk part of the integral (\ref{hsig}) can be given as:
$$\tau_{bulk} = - \frac1{24 \pi^2} \int_M \text{Tr} \ R \wedge R = - \frac1{24 \pi^2} \int_M R_{ab} \wedge R^{ab}$$
However, we can see from (\ref{rmatrix}) that every element of the curvature 2-forms has a single 2-form 
term. Thus, we can write:
$$R_{ab} \wedge R^{ab} = 0 \qquad \Rightarrow \qquad \tau_{bulk} = 0$$
Now, as in the case of $\chi(M)$, the boundary integral term also vanishes following the same logic. 
$$\theta_{ab} \wedge R^{ab} = 0 \qquad \Rightarrow \qquad \tau_{boundary} = 0$$
This leaves us with nothing but the last term, known as the spectral asymmetry term $\eta_S ( \partial M )$ 
which in this case is also known to vanish. Therefore:
\begin{equation}
\tau (M) = 0
\end{equation}
As before, analyzing from the point of view of $SU(2)_\pm$ gauge fields lets us use:
$$\delta_{\mu \lambda} \delta_{\nu \gamma} - \delta_{\mu \gamma} \delta_{\nu \lambda} = \frac12 \Big( \eta^{(+)i}_{\mu \nu} \eta^{(+)i}_{\lambda \gamma} + \eta^{(-)i}_{\mu \nu} \eta^{(-)i}_{\lambda \gamma} \Big)$$
to write the bulk part of the signature complex as
$$\tau_{bulk} = - \frac1{24 \pi^2} \int_M Tr \big( R \wedge R \big) = - \frac23 \Big( \chi^+_{bulk} - \chi^-_{bulk} \Big) = \frac23 \big( - 1 + 1 \big) = 0$$
where we can see that the individual bulk contribution is:
\begin{equation}
\begin{split}
&\tau_{bulk} = \tau^+_{bulk} + \tau^-_{bulk}\\ 
\tau^+_{bulk} = - \frac23 \chi^+_{bulk} &= - \frac23 \qquad \tau^-_{bulk} = \frac23 \chi^-_{bulk} = \frac23
\end{split}
\end{equation}
which concludes our computation of topological invariants of the Euclidean Schwarzschild metric.

\section{Discussion}

In this note we have started applying the bottom-up approach of emergent gravity to (Euclidean)
Schwarzschild solution which we dub as emergent Schwarzschild. The emergent 
Schwarzschild solution describes a Ricci-flat manifold, although it is not a K\"ahler manifold. 
So it does not admit a natural symplectic structure. The best alternative choice as was  utilized by 
Etesi and Hausel \cite{bh-harmonic}) was to consider the (anti) self-dual harmonic two-forms on 
the space and define a Poisson algebra determined by the self-dual harmonic two-form. However 
a magnetic mass (and an electric mass) at the origin seems to violate the Jacobi identity of the 
underlying Poisson algebra which can be circumvented going to Euclidean signature and using Kruskal Szekeres 
coordinates. Therefore the Schwarzschild instanton always remained a challenging goal 
to pursue from the bottom-up approaches of emergent gravity. 

We have found a suitable Darboux chart for the emergent Schwarzschild solution
for which locally we have the Jacobi identity satisfied for the symplectic $U(1)$ gauge fields 
emergent from the metric as well as  the Bianchi identity for the vector fields. We set up the 
Seiberg Witten map between the commutative and non-commutative description and did a 
thorough geometrical engineering for the instanton solution. We saw that the two
instantons forming the emergent Schwarzschild solution belong to different gauge groups namely 
$SU(2)_{L}$ and $SU(2)_{R}$ and hence they can't decay into a vacuum thus explaining the 
stability of emergent Schwarzschild space against perturbation, which might be a generic fact for 
any Ricci-flat four manifold as ours. The emergent nature of Taub-NUT instanton
and its connection with dynamical systems have been discussed in (\cite{TNdynamics}). 

It will be interesting to investigate how to analyze a charged black hole solution in this bottom-up 
approach of emergent gravity. In \cite{Wang:1998kb}, it was suggested that there are two
kinds of 4D and 2D EBHs in nature. The first kind of extremal black holes can be obtained by 
first taking the extreme limit and then the boundary limit starting from general non-extremal 
configuration. The entropy of this kind of EBH is zero. The second kind of EBH which still holds the 
topological configuration of NEBH can be obtained by taking the boundary limit first and then the
extreme limit. This kind of EBH satisfies the BH entropy formula. These two kinds of EBHs 
have different intrinsic thermodynamical properties owing to different topological characteristics 
playing an essential role in the classification of these solutions. For the first kind, the Euler 
characteristic is zero; and for the second, it is equal to two or one provided they are 4D or 2D 
EBHs respectively. Now it will be interesting to address the fact whether such a change in 
topology of spacetime can be explained from the point of view of a well-defined mechanism 
inspired by the emergent gravity  approach which was set up by one of the authors in 
\cite{Lee:2012ju}.

\begin{flushleft}
\textbf{Acknowledgments: } The research of RR was supported by FAPESP through Instituto de 
Fisica, Universidade de Sao Paulo with grant number 2013/17765-0. This work was performed
during RR's visit to S.N. Bose National Centre for Basic Sciences in Kolkata. He thanks SNBNCBS 
for the hospitality and support during that period.
\end{flushleft}

\section{APPENDIX}

\subsection{Relations from Differential Geometry}

The Cartan structure equations are powerful tools in differential geometry, useful for the analysis 
of curvature involved in General Relativity. Cartan's first structure equation is:
\begin{equation}
T^a = d e^a + {\omega^a}_b \wedge e^b
\end{equation}
Under torsion free condition ($T^a = 0$), we have:
$$d e^a = - {\omega^a}_b \wedge e^b \quad \Rightarrow \quad \partial_\mu {e^a}_\nu = - {{\omega_\mu}^a}_b {e^b}_\nu$$
Upon contraction with ${E_c}^\nu$, we can proceed to write:
$${E_c}^\nu \partial_\mu {e^a}_\nu = - {{\omega_\mu}^a}_b \big( {e^b}_\nu {E_c}^\nu \big) \qquad \Rightarrow \qquad \underbrace{\partial_\mu \big( {E_c}^\nu {e^a}_\nu \big)}_0 - {e^a}_\nu \partial_\mu {E_c}^\nu = - {{\omega_\mu}^a}_b \delta^b_c = - {{\omega_\mu}^a}_c$$
\begin{equation}
\therefore \quad \partial_\mu {E_c}^\nu = {{\omega_\mu}^a}_c {E_a}^\nu
\end{equation}

\subsection{Permutation operations with the t'Hooft symbols}

Since each of the t'Hooft symbols has only one non-zero element in each row and column, they are 
permutation matrices. Here, we will just establish the permutation rules associated with each of 
the matrices. The t'Hooft symbols are given by:
\begin{align}
\eta^{(+)1} = \left({\begin{array}{cccc}
0 & 0 & 0 & 1\\
0 & 0 & 1 & 0\\
0 & -1 & 0 & 0\\
-1 & 0 & 0 & 0
\end{array} } \right) \quad
\eta^{(+)2} = \left({\begin{array}{cccc}
0 & 0 & -1 & 0\\
0 & 0 & 0 & 1\\
1 & 0 & 0 & 0\\
0 & -1 & 0 & 0
\end{array} } \right) \quad
\eta^{(+)3} = \left({\begin{array}{cccc}
0 & 1 & 0 & 0\\
-1 & 0 & 0 & 0\\
0 & 0 & 0 & 1\\
0 & 0 & -1 & 0
\end{array} } \right) \\ 
\eta^{(-)1} = \left({\begin{array}{cccc}
0 & 0 & 0 & -1\\
0 & 0 & 1 & 0\\
0 & -1 & 0 & 0\\
1 & 0 & 0 & 0
\end{array} } \right) \quad
\eta^{(-)2} = \left({\begin{array}{cccc}
0 & 0 & -1 & 0\\
0 & 0 & 0 & -1\\
1 & 0 & 0 & 0\\
0 & 1 & 0 & 0
\end{array} } \right) \quad
\eta^{(-)3} = \left({\begin{array}{cccc}
0 & 1 & 0 & 0\\
-1 & 0 & 0 & 0\\
0 & 0 & 0 & -1\\
0 & 0 & 1 & 0
\end{array} } \right)
\end{align}
Suppose the arrangement with the original matrix in terms of rows is 1234. Then, left-multiplication 
with each of the t'Hooft symbols gives the following:

\begin{center}

\textbf{t'Hooft Permutations}\\

\begin{tabular}{|c|c|c|c|}

\hline & & &\\

\textbf{Symbol} & $\bm{\eta^1}$  &  $\bm{\eta^2}$ & $\bm{\eta^3}$\\
\hline & & &\\

$SU(2)_+$
&
$43 \bar{2} \bar{1}$
&
$34 \bar{1} \bar{2}$
&
$2 \bar{1} 4 \bar{3}$ \\

\hline & & &\\

$SU(2)_-$
&
$\bar{4} 3 \bar{2} 1$
&
$3 \bar{4} \bar{1} 2$
&
$2 \bar{1} \bar{4} 3$ \\

\hline
\end{tabular}
\end{center}

where numbers labelled as $\bar{X}$ are rows where the sign has been flipped.
\subsection{$\bold{SU(2)_\pm}$ gauge fields}
The various gauge fields involved with equation (\ref{gf}) are given by:
\begin{align}
A^{(+)1} &= 
- \frac1{4r} Tr \left({\begin{array}{cccc}
0 &  b \ e^4 & c \ e^4 & 0\\
0 &  b \ e^3 & 0 & - c \ e^4\\
a \ e^1 & 0 & b \ e^3 & b \ e^4\\
0 & - a \ e^1 & 0 & 0
\end{array} } \right) \quad
A^{(-)1} = 
- \frac1{4r} Tr \left({\begin{array}{cccc}
0 &  - b \ e^4 & - c \ e^4 & 0\\
0 &  b \ e^3 & 0 & - c \ e^4\\
a \ e^1 & 0 & b \ e^3 & b \ e^4\\
0 & a \ e^1 & 0 & 0
\end{array} } \right) \nonumber \\
&= - \frac1{2r} b \ e^3 \hspace{7cm} = - \frac1{2r} b \ e^3\\
A^{(+)2} &= 
- \frac1{4r} Tr \left({\begin{array}{cccc}
0 &  - b \ e^3 & 0 & c \ e^4\\
0 &  b \ e^4 & c \ e^4 & 0\\
0 & a \ e^1 & 0 & 0\\
a \ e^1 & 0 & b \ e^3 & b \ e^4
\end{array} } \right) \qquad
A^{(-)2} = 
- \frac1{4r} Tr \left({\begin{array}{cccc}
0 &  - b \ e^3 & 0 & c \ e^4\\
0 & - b \ e^4 & - c \ e^4 & 0\\
0 & a \ e^1 & 0 & 0\\
- a \ e^1 & 0 & - b \ e^3 & - b \ e^4
\end{array} } \right) \nonumber \\
&= - \frac1{2r} b \ e^4 \hspace{7.1cm} = - \frac1{2r} b \ e^4\\
A^{(+)3} &= 
- \frac1{4r} Tr \left({\begin{array}{cccc}
- a \ e^1 & 0 & - b \ e^3 & - b \ e^4\\
0 & - a \ e^1 & 0 & 0\\
0 &  b \ e^4 & c \ e^4 & 0\\
0 & - b \ e^3 & 0 & c \ e^4\\
\end{array} } \right) \hspace{0.7cm}
A^{(-)3} = 
- \frac1{4r} Tr \left({\begin{array}{cccc}
- a \ e^1 & 0 & - b \ e^3 & - b \ e^4\\
0 & - a \ e^1 & 0 & 0\\
0 & - b \ e^4 & - c \ e^4 & 0\\
0 & b \ e^3 & 0 & - c \ e^4\\
\end{array} } \right) \nonumber \\
&= \frac1{2r} \big( a \ e^1 - c \ e^4 \big) \hspace{6.5cm} = \frac1{2r} \big( a \ e^1 + c \ e^4 \big)
\end{align}
\begin{align}
\label{gf1} F^{(+)1} &= 
- \frac{m}{4r^3} Tr \left({\begin{array}{cccc}
z & - \bar{y} & - 2 \bar{x} & 0\\
y & \bar{z} & 0 & 2 \bar{x}\\
2 x & 0 & \bar{z} & - \bar{y}\\
0 & - 2 x & y & z
\end{array} } \right) \qquad
F^{(-)1} = 
- \frac{m}{4r^3} Tr \left({\begin{array}{cccc}
- z & \bar{y} & 2 \bar{x} & 0\\
y & \bar{z} & 0 & 2 \bar{x}\\
2 x & 0 & \bar{z} & \bar{y}\\
0 & 2 x & - y & - z
\end{array} } \right) \nonumber \\ \displaybreak[0]
&= - \frac{m}{2r^3} \big( z + \bar{z} \big) \hspace{5.9cm} = \frac{m}{2r^3} \big( z - \bar{z} \big)\\
\label{gf2} F^{(+)2} &= 
- \frac{m}{4r^3} Tr \left({\begin{array}{cccc}
- y & - \bar{z} & 0 & - 2 \bar{x}\\
z & - \bar{y} & - 2 \bar{x} & 0\\
0 & 2 x & - y & - z\\
2 x & 0 & \bar{z} & - \bar{y}
\end{array} } \right) \qquad
F^{(-)2} = 
- \frac{m}{4r^3} Tr \left({\begin{array}{cccc}
- y & - \bar{z} & 0 & - 2 \bar{x}\\
- z & \bar{y} & 2 \bar{x} & 0\\
0 & 2 x & - y & - z\\
- 2 x & 0 & - \bar{z} & \bar{y}
\end{array} } \right) \nonumber \\
&= \frac{m}{2r^3} \big( y + \bar{y} \big) \hspace{6.4cm} = \frac{m}{2r^3} \big( y - \bar{y} \big)\\
\label{gf3} F^{(+)3} &= 
- \frac{m}{4r^3} Tr \left({\begin{array}{cccc}
- 2 x & 0 & - \bar{z} & \bar{y}\\
0 & - 2 x & y & z\\
z & - \bar{y} & - 2 \bar{x} & 0\\
- y & - \bar{z} & 0 & - 2 \bar{x}
\end{array} } \right) \hspace{0.7cm}
F^{(-)3} = 
- \frac{m}{4r^3} Tr \left({\begin{array}{cccc}
- 2 x & 0 & - \bar{z} & - \bar{y}\\
0 & - 2 x & y & z\\
- z & \bar{y} & 2 \bar{x} & 0\\
y & \bar{z} & 0 & 2 \bar{x}
\end{array} } \right) \nonumber \\
&= \frac{m}{r^3} \big( x + \bar{x} \big) \hspace{6.65cm} = \frac{m}{r^3} \big( x - \bar{x} \big)
\end{align}

\end{document}